\documentclass{PoS}
\usepackage{graphicx, epsf}
\usepackage{epsfig}

\title{Searching for a continuum 4D field theory arising from a 5D non-abelian gauge theory}

\ShortTitle{Searching for a continuum 4D field theory arising from a 5D non-abelian gauge theory}

\author{Luigi Del Debbio, Richard D. Kenway, {\speaker{Eliana Lambrou}} and Enrico Rinaldi\\
       The Higgs Centre for Theoretical Physics, School of Physics and Astronomy,\\ University of Edinburgh\\
       Edinburgh, EH9 3JZ, UK\\
       E-mail: \email{luigi.del.debbio@ed.ac.uk, r.d.kenway@ed.ac.uk, e.lambrou@ed.ac.uk, e.rinaldi@ed.ac.uk}}

\abstract{The anisotropic 5D SU(2) Yang-Mills model has been widely investigated on the lattice during the last decade. In the case where all dimensions are large in size, it was previously claimed that there is a new phase in the phase diagram, called the Layer phase. In this phase, the gauge fields would be localized on 4D layers. Previous works claim that the phase transition to the Layer phase is of second order, which would allow a continuum limit to be taken. We present the extension of the previous work to large lattices, for which we found a first order phase transition. This leaves the scenario that this 5D theory can be dimensionally reduced to a continuum 4D field theory, doubtful. \\

\begin{flushright}
Edinburgh 2013/24 \\
\end{flushright}
}

\FullConference{31st International Symposium on Lattice Field Theory LATTICE 2013\\
		 July 29 - August 3, 2013\\
		 Mainz, Germany}

\begin{document}

\section{Introduction}
Models with extra dimensions have been developed in order to investigate the gauge hierarchy problem, the cosmological constant problem and the fermion mass hierarchy problem. In all of these, the main requirement is to find a mechanism which dimensionally reduces the model to the usual four-dimensional spacetime that we live in. This is achieved by compactification or localization. The latter is based on the brane world scenario which visualises the observed world as a four dimensional hyperplane embedded in the bulk. Many models were developed based on this scenario, out of which the most well-known are the Randall-Sundrum~(RS) and the Dvali-Shifman~(DS) models. In these, one of the main problems is to achieve localization of the gauge fields on the branes. 

In the early 1980s, Fu and Nielsen introduced the idea that one can achieve gauge field localization by imposing an anisotropy between the interactions in the usual four-dimensional spacetime and the extra dimension~\cite{Fu}. This anisotropy gives rise to a new phase, called the Layer phase, in which the fields are free to propagate on the Layers (in the usual four dimensions) but are confined in the extra dimension. So, one can think of the five-dimensional world as being reduced to many non-interacting copies of four-dimensional layers embedded in an extra dimension.  

This idea of the dimensional reduction of a system via localization in the presence of this Layer phase has already been investigated for abelian and non-abelian groups. In the latter case, the five-dimensional SU(2) Yang-Mills model was investigated using both the Mean-Field approximation~\cite{MeanField1} and Monte Carlo numerical simulations~\cite{Farakos, Murata}. All works seem to agree that, when the lattice spacing in the usual four dimensions is smaller than the lattice spacing in the extra dimension, the standard bulk phase transition that separates the confined from the deconfined phase in the five-dimensional non-abelian gauge theory, becomes a second order phase transition. If this is the case then one expects to be able to take a continuum limit, in the sense of the presence of a non-trivial fixed point and thus be able to define a continuum four-dimensional field theory.      

In this work, we present a summary of the extension of the previous Monte Carlo numerical simulations to larger volumes to check the validity of the claim that there is a second order phase transition into the Layer phase in the non-abelian pure gauge theory~\cite{Eliana}. 
 
\section{Lattice set-up}
In this work, our interest  focuses on the five-dimensional non-abelian pure gauge theory. The exploration of this system, as mentioned in the introduction, requires an anisotropy between the usual four dimensions and the extra dimension and, as a consequence, different lattice spacing in the usual four dimensions ($a_4$) and the extra dimension ($a_5$). This is achieved by imposing different lattice couplings in the Wilson action for the plaquettes along space-time directions ($\beta_4$) and along the extra dimension ($\beta_5$). Therefore, the Wilson action of the model is given by
\begin{equation}
	S_W=\beta_4\sum_x \sum_{1\leq\mu < \nu\leq 4}\Big( 1 - \frac{1}{N_c}\mbox{Tr} U_{\mu\nu}(x)  \Big) + \beta_5\sum_x \sum_{1\leq\mu\leq 4}\Big( 1- \frac{1}{N_c}\mbox{Tr} U_{\mu 5}(x) \Big)
\end{equation}
where the space-time plaquette is given by
\begin{equation}
	U_{\mu\nu}=U_\mu(x) U_\nu(x+\hat\mu a_4) U^{\dagger}_\mu (x+\hat \nu a_4)U^{\dagger}_\nu(x)
\end{equation}
and the extra-dimensional plaquette is given by
\begin{equation}
	U_{\mu 5}=U_\mu(x) U_5(x+\hat\mu a_4) U^{\dagger}_\mu (x+\hat 5 a_5)U^{\dagger}_5(x).
\end{equation}
In order to recover the non-abelian pure gauge action in the continuum limit, we define the gauge links to be
\begin{equation}
U_{\mu} = \exp(ig_5a_4A_\mu) \;\;\;\;\;\;\;\; \mbox{and} \;\;\;\;\;\;\;\; U_5=\exp(ig_5a_5A_5)
\end{equation}
and we recognise the relations between the lattice couplings $\beta_4$, $\beta_5$ and the lattice spacings $a_4$, $a_5$ to be
\begin{equation}
\beta_4= \frac{2N_c a_5}{g_5^2}\;\;\;\;\;\;\;\; \mbox{and} \;\;\; \;\;\;\;\; \beta_5=\frac{2N_ca_4^2}{a_5g_5^2}
\end{equation} 
where $g_5$ is the gauge coupling in the five-dimensional Yang-Mills theory and $N_c$ is the number of colours, which in this work is set to 2.
\\
The anisotropy parameter is given by 
\begin{equation}
\gamma = \sqrt{\frac{\beta_5}{\beta_4}}
\end{equation}
and at tree-level it becomes
\begin{equation}
	\gamma=\frac{a_4}{a_5}	
\end{equation}
\\
The main observables that we use are the average of the extra-dimensional plaquette, $\hat P_5$
\begin{equation}
	\langle \hat P_5 \rangle = \Big \langle \frac{1}{8V}\sum_x \sum_\mu \mbox{Tr}(U_\mu 5(x)) \Big \rangle,
\end{equation}
the Polyakov loop in the temporal direction
\begin{equation}
	Poly_T = \frac{L_T}{2V} \bigg | \sum_{\vec{x}, x_5} \mbox{Tr} \prod_{x_1=0}^{(L_T-1)a_4}U_1(x) \bigg |
\end{equation}
and similarly, with the appropriate changes, the Polyakov loop along the extra dimension, $Poly_5$.  Using the appropriate analysis, their susceptibilities and distributions are extracted.

\section{Results} 
\begin{figure}[!h] 
	\centering
	\includegraphics[angle=270,scale=0.5]{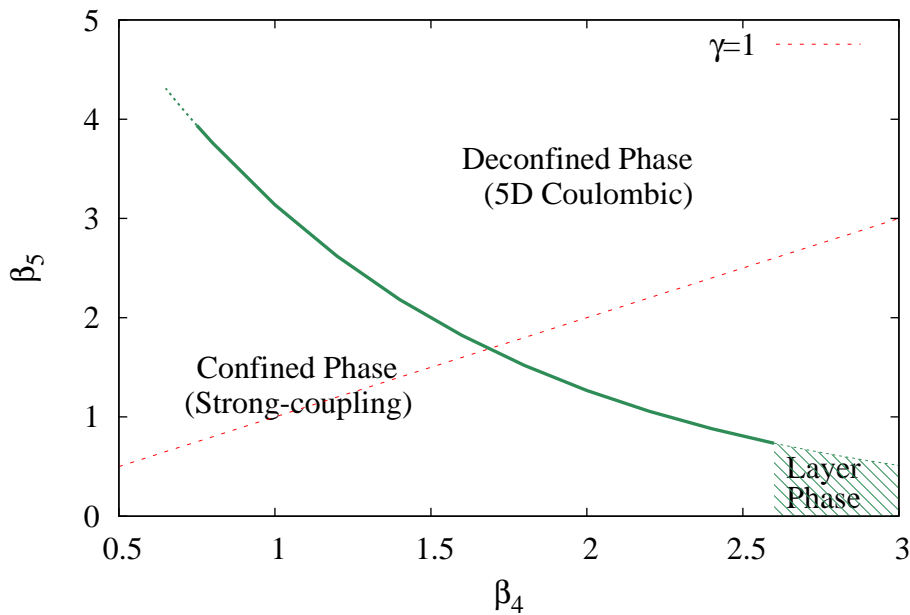}
	\caption{A cartoon of the phase diagram of the five-dimensional model in the infinite volume limit. The green solid line shows a first order transition between the strong-coupling and the Coulombic phases. In the shaded region, where $\beta_4 \geq 2.60$, it was previously claimed that the order of the phase transition changes to second and the Layer phase can be defined.}
	\label{fig:PhaseDiagram}
\end{figure}

The phase diagram of the model described above was investigated focusing specifically on the region where the Layer phase is believed to exist. A schematic phase diagram of the above model from combining our results with those from previous work~\cite{Ejiri,Forcrand,Francesco,ER}, is shown in Fig.~\ref{fig:PhaseDiagram}. The regime of our interest is at $\gamma < 1$, where the Layer phase is believed to exist, i.e. in the shaded region of Fig.~\ref{fig:PhaseDiagram}. In ~\cite{Francesco}, it was shown that a bulk phase transition is present up to $\beta_4=2.50$ (blue solid line in Fig.\ref{fig:PhaseDiagram}) and as one moves towards larger values of $\beta_4$, bigger volumes are required in order to see this first order phase transition. Specifically for $\beta_4=2.50$ at least 20 spatial/temporal points were required to see this transition. 

Farakos and Vrentzos claimed that the order of the phase transition changes to second for $\beta_4\geq 2.60$, but the largest volume used to explore this was $16^5$, which is smaller than the minimum size that is required according to Knechtli et al.~\cite{Francesco}. Therefore, we decided to explore this further to see if the second order phase transition is still present for larger volumes.        

The procedure that we followed was to keep the value of $\beta_4$ constant and vary $\beta_5$ to find the critical point. The choice of $\beta_4$ was based on the previous work of~\cite{Farakos} and was set to $\beta_4=2.60$. Understanding the importance of the extrapolation to the thermodynamic limit, we implemented the model on larger volumes of $20^4\times8$ and $24^4\times8$ but we also took measurements for $V=16^5$ to check the consistency of our results with~\cite{Farakos}. Since the lattice spacing in the extra dimension is much larger than the lattice spacing in the usual four dimensions, reducing the lattice points along the extra dimension does not introduce any significant finite-size effect and it was checked that the extra-dimensional Polyakov loop was unbroken, ensuring that the system is still in the infinite volume limit. The configurations were obtained by applying a combination of Kennedy-Pendleton Heat-Bath algorithm with overrelaxation steps and the above observables where measured for an ensemble of 100,000 - 200,000 of these configurations.  For each set of points, ($\beta_4, \beta_5$), measurements were taken starting from both cold  (unity) and hot (random) configurations. 

\begin{figure}[!h]
	\centering
	\includegraphics[angle=270,scale=0.4]{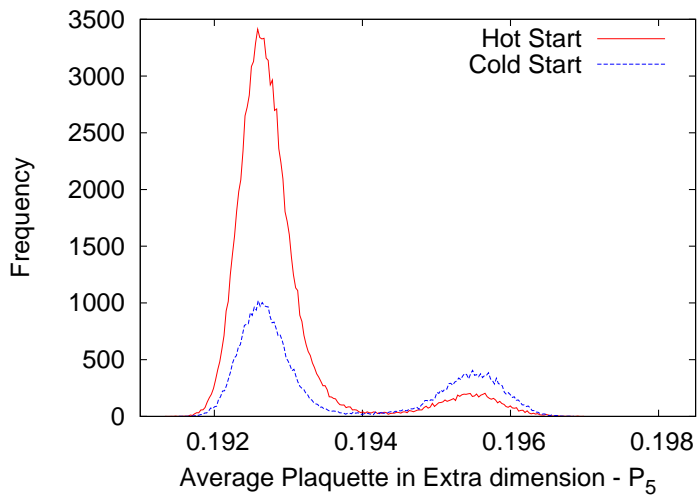}
	\caption{The number distributions of the extra-dimensional plaquette starting from both oriented and random configurations for the parameter point $(2.60,0.8435)$ and $V=24^4\times8$. A two-peak structure is apparent which indicates a first-order phase transition.}
	\label{fig:Plaq_24}
\end{figure}

First, the critical point for our smallest volume $V=16^5$ was found by applying reweighting techniques to the susceptibility of the extra-dimensional plaquette and was estimated to be $\beta_{5c}=0.8437(5)$, which matches the value in~\cite{Farakos}. However, when we moved to $V=20^4\times8$ the fluctuations of the extra-dimensional plaquette close to the critical point were large, signalling that there might be a first order phase transition, so we took measurements for $V=24^4\times 8$. For the latter volume, there was a clear two-peak structure with a minor fluctuation between the two vacua, as it is shown in Fig.~\ref{fig:Plaq_24}, which indicates a first order phase transition. For completeness, we also implemented one point close to the critical one ($\beta_5=0.844)$ for $V=32^4\times 8$, where the extra-dimensional plaquette shows a clear two-state signal with no fluctuation between the vacua. So we conclude that the transition is of first order.

From the above results, it is clear that the bulk phase transition between the confined and deconfined phase continues up to $\beta_4=2.60$ with no indication that for larger values of $\beta_4$ it will change to a second order phase transition to give the possibility of defining a continuum four-dimensional field theory. Although the change of the order of the phase transition to a second order cannot be excluded, an extension of the investigation should be done very carefully, especially in the choice of lattice volumes in order not to get misleading results.

\section{Conclusions}
This work shows the extension of the exploration of the transition to a layered phase in the five-dimensional anisotropic SU(2) Yang-Mills model using Monte Carlo simulations. This is done in the regime where the lattice spacing in the usual four dimensions is smaller than in the extra dimension, i.e. the regime where $\gamma < 1$. This transition was previously claimed to be second order for lattice couplings $\beta_4 \geq 2.60$ and thus the five-dimensional theory could be dimensionally reduced to a continuum four-dimensional theory; this implies that  in the Layer phase a continuum effective 4D theory could be defined. However, our lattice simulations give a clear indication of a first order transition at $\beta_4=2.60$, which implies that up to this value there is no evidence of this dimensional reduction. We cannot exclude the possibility that the transition will change from first to second order for higher values of $\beta_4$, but the requirement to go to much larger lattices in order to unambiguously determine the order of the transition makes the continuation of this search numerically very demanding.

\end{document}